\documentclass[12pt,onecolumn,technote]{IEEEtran}

\usepackage{mathrsfs}
\usepackage{txfonts}
\usepackage [dvips] {graphicx}

\ifCLASSINFOpdf

\else

\fi

\renewcommand\baselinestretch{2.0}

\begin{document}

\title{Sidelobe Suppression for Robust Beamformer via The Mixed Norm Constraint}

\author{Yipeng~Liu,
        and~Qun~Wan
\thanks{Yipeng Liu and Qun Wan are with the Electronic Engineering Department,
University of Electronic Science and Technology of China, Chengdu, 611731 China
e-mail: ({liuyipeng,~wanqun}@uestc.edu.cn).}
\thanks{Manuscript received Month Day, 2010; revised Month Day, Year.}}

\markboth{Journal Title,~Vol.~X, No.~X, Month~Year}%
{Shell \MakeLowercase{\textit{et al.}}: Bare Demo of IEEEtran.cls for Journals}

\maketitle

\begin{abstract}

Applying a sparse constraint on the beam pattern has been suggested
to suppress the sidelobe of the minimum variance distortionless
response (MVDR) beamformer recently. To further improve the
performance, we add a mixed norm constraint on the beam pattern. It
matches the beam pattern better and encourages dense distribution in
mainlobe and sparse distribution in sidelobe. The obtained
beamformer has a lower sidelobe level and deeper nulls for
interference avoidance than the standard sparse constraint based
beamformer. Simulation demonstrates that the SINR gain is
considerable for its lower sidelobe level and deeper nulling for
interference, while the robustness against the mismatch between the
steering angle and the direction of arrival (DOA) of the desired
signal, caused by imperfect estimation of DOA, is maintained too.
\end{abstract}


\IEEEpeerreviewmaketitle

\section{Introduction}

\IEEEPARstart{M}{ultiple-antenna} systems have received a lot of
attention from both the wireless industry and academia, because of
their strong potential in realizing high date rate wireless
communications in next generation wireless networks. A beamformer is
a versatile form of spatial filtering. It uses multiple antenna
systems to separate signals that have overlapping frequency spectra
but originate from different spatial locations. Beamforming has
become a key technique in current and future wireless communications
\cite{rab}.

The minimum variance distortionless response (MVDR) beamformer has
been considered as a popular method for enhancing the signal from
the desired direction while suppressing all signals from other
directions as well as the background noise \cite{rab}, but its
relatively high sidelobe level would lead to significant performance
degradation, especially with the unexpected increase in interference
or background noise \cite{pamvdr}. To enhance the robustness in the
presence of array steering vector errors, doubly constrained robust
capon beamformer used a norm constraint on the weight vector to
improve the robustness \cite{dcrcb}. To achieve a faster convergence
speed and a higher steady state signal to interference plus noise
ratio (SINR) \cite{lqc} constrains its weight vector to a specific
conjugate symmetric form. In \cite{rbf}, fully complex-valued radial
basis function (RBF) network with the fully complex-valued
activation function is used in MVDR beamformer to short the
convergence period. However, high sidelobe is another drawback,
which would result in deep degradations in the case of unexpected
interferences or an increase in noise power \cite{pamvdr}. In order
to provide sidelobe suppression and angle mistmatch robustness for
an MVDR beamformer, a sparse constraint on the whole beam pattern
was recently proposed in \cite{sbf}, and a robust beamformer with
the sidelobe suppressed was obtained.

In \cite{sbf}, the sparse constraint was added equally on both the
mainlobe and the sidelobe. But the expected beam pattern would enjoy
most of the high array gains in the mainlobe, and the array gains in
the mainlobe are in dense distribution. The constraint should
encourage dense distribution in mainlobe and sparse distribution in
sidelobe. To further enhance the performance, a mixed norm
constraint is incorporated to match the expected beam pattern
better. Numerical evaluations show that the proposed beamformer
achieves a lower sidelobe level, deeper nulls for interference
avoidance.

\section{MVDR Beamformer}

The signal received by a uniform linear array (ULA) with \emph{M}
antennas can be represented by an \emph{M}-by-1 vector,
\textbf{x}(\emph{k}), the expression of which is given by

\begin{equation}
\label{eq1} {\bf{x}}(k) = s(k){\bf{a}}(\theta _0 ) + \sum\limits_{j
= 1}^J {\beta _j (k){\bf{a}}(\theta _j )}  + {\bf{n}}(k) \label{eq2}
\end{equation}
where \emph{k} is the index of time, \emph{J} is the number of
interference sources, s(\emph{k}) and $\beta _i (k)$ (for \emph{j} =
1,~...~,~ \emph{J}) are the amplitudes of the SOI (signal of
interest) and interfering signals at time instant \emph{k},
respectively, $\theta _l$ (for \emph{l} = 0, 1,~...~, \emph{J}) are
the DOAs of the SOI and interfering signals, $ \varphi _l  = \left(
{{{2\pi d} \mathord{\left/
 {\vphantom {{2\pi d} \lambda }} \right.
 \kern-\nulldelimiterspace} \lambda }} \right)\sin \theta _l
$, with \emph{d} being the distance between two adjacent antennas
and $\lambda$ being the operating wavelength \cite{rab}, i.e., the
wavelength of the SOI, and \textbf{n}(\emph{k}) is the additive
white Gaussian noise (AWGN) vector at time instant \emph{k}; $
{\bf{a}}(\theta ) $ is the steering vector in the angle $ \theta $,
with its \emph{m}-th element $ \exp \left( {j(m - 1)\frac{{2\pi
d}}{\lambda }\sin \theta } \right) $.

The output of a beamformer for the time instant \emph{k} is then
given by

\begin{equation}
\label{eq2} y(k) = {\bf{w}}^H {\bf{x}}(k) = s(k){\bf{w}}^H
{\bf{a}}(\theta _0 ) + \sum\limits_{j = 1}^J {\beta _j (k){\bf{w}}^H
{\bf{a}}(\theta _j )}  + {\bf{w}}^H {\bf{n}}(k)
\end{equation}
where \textbf{w} is the \emph{M}-by-1 complex-valued weighting
vector of the beamformer.

The MVDR beamformer is designed to minimize the total array output
energy, subject to a linear distortionless constraint on the SOI.
The weighting vector of the MVDR beamformer \cite{rab} is given by

\begin{equation}
\label{eq3} {\bf{w}}_{MVDR}  = \mathop {\arg \min }\limits_{\bf{w}}
\left( {{\bf{w}}^H {\bf{R}}_x {\bf{w}}} \right),{\rm{
s}}{\rm{.t}}{\rm{. }}{\bf{w}}^H {\bf{a}}(\theta _0 ) = 1
\end{equation}
where ${{\bf{R}}_x }$  is the \emph{M}-by-\emph{M} covariance matrix
of the received signal vector \textbf{x}(\emph{k}), and ${\bf{w}}^H
{\bf{a}}(\theta _0 ) = 1 $ is the distortionless constraint applied
on the SOI.

\section{The Mixed Norm Constraint Beamformer}

In order to suppress the sidelobe level of the conventional MVDR
beamformer, a sparse constraint on the whole beam pattern was
suggested in \cite{sbf}. Accordingly, the weighting vector of the
improved MVDR beamformer based on a sparse constraint (SC) is given
by

\begin{equation}
\label{eq4} {\bf{w}}_{SC}  = \mathop {\arg \min }\limits_{\bf{w}}
\left( {{\bf{w}}^H {\bf{R}}_x {\bf{w}} + \gamma _1 \left\|
{{\bf{w}}^H {\bf{A}}} \right\|_p^p } \right),{\rm{
s}}{\rm{.t}}{\rm{. }}~{\bf{w}}^H {\bf{a}}(\theta _0 ) = 1
\end{equation}
where  $ \gamma _{\rm{1}} $ is the factor that controls the tradeoff
between the minimum variance constraint on the total array output
energy and the sparse constraint on the beam pattern, The M-by-N
matrix \textbf{A} is the steering matrix with $ \alpha _n $s ( n =
1, 2, ... , N ) being the sampled angles in the [ -90$ ^ \circ $,
90$ ^ \circ $], and it covers all the N steering vectors for all
possible interference with DOA in the sampling range, with $ \alpha
_0 $ being the DOA of the SOI as defined in (\ref{eq1}), i.e.
\begin{equation}
\label{eq5} {\bf{A}} = \left[ {\begin{array}{*{20}c}
   1 &  \cdots  & 1  \\
   {\exp \left( {j\varphi _1 } \right)} &  \cdots  & {\exp \left( {j\varphi _N } \right)}  \\
    \vdots  &  \ddots  &  \vdots   \\
   {\exp \left( {j\left( {M - 1} \right)\varphi _1 } \right)} &  \cdots  & {\exp \left( {j\left( {M - 1} \right)\varphi _N } \right)}  \\
\end{array}} \right]
\end{equation}
\begin{equation}
\label{eq6} \varphi _n  = \frac{{2\pi d}}{\lambda }\sin \alpha _n
,~{\rm{ for }}~n = 1, \cdots ,N
\end{equation}
and $\left\| {\bf{x}} \right\|_p  = \left( {\sum\nolimits_i {\left|
{x_i } \right|^p } } \right)^{1/p}$ is is the $\mathscr{C}_p$\ norm
of a vector \textbf{x}.When $0 \le p \le 1 $, the $\mathscr{C}_p$\
norm provides a measurement of sparsity for \textbf{x}. The smaller
the value of $\left\| {\bf{x}} \right\|_p^p $ is. the sparser the
vector \textbf{x} is, It means that the number of trivial entries in
x is larger \cite{sbf}. When p = 1, (\ref{eq4}) is an second order
cone programming (SOCP), and can be solved efficiently.

The optimal weighting vector indicated by (\ref{eq4}) can be found
by an adaptive iteration algorithm \cite{sbf,bdrao}. When \emph{p} =
1, a simpler way called basis pursuit \cite{bp}, can solve
(\ref{eq4}) efficiently.

In (\ref{eq4}), the sparse constraint operates for all the array
gains $ {\bf{w}}^H {\bf{A}} $ in all the possible values of DOA from
-90$ ^ \circ $ to 90$ ^ \circ $, i.e. it enforces sparse
distribution of the array gains in both the mainlobe and the
sidelobe. However, the array gains are not in standard sparse
distribution, but in dense distribution in the mainlobe and in
sparse distribution in the sidelobe. Here instead of the standard
sparse constraint, an mixed norm constraint with different norms on
different lobes can be added to the MVDR beamformer to improve the
performance. It can be formulated as

\begin{equation}
\label{eq7}
\begin{array}{c}
 {\bf{w}}_{MNB}  = \mathop {\arg \min }\limits_{\bf{w}} \left[ {{\bf{w}}^H {\bf{R}}_x {\bf{w}} + \gamma _2 \left( {\left\| {{\bf{w}}^H {\bf{A}}_M } \right\|_\infty  + \left\| {{\bf{w}}^H {\bf{A}}_S } \right\|_1^{} } \right)} \right] \\
 {\rm{  s}}{\rm{.t}}{\rm{.  }}{\bf{w}}^H {\bf{a}}(\theta _0 ) = 1 \\
 \end{array}
\end{equation}
where

\begin{equation}
\label{eq8} {\bf{A}}_M  = \left[ {\begin{array}{*{20}c}
   {{\bf{a}}\left( {\theta _{ - b} } \right)} &  \cdots  & {{\bf{a}}(\theta _0 )} &  \cdots  & {{\bf{a}}(\theta _{ + b} )}  \\
\end{array}} \right]
\end{equation}

\begin{equation}
\label{eq9} {\bf{A}}_S  = \left[ {\begin{array}{*{20}c}
   {{\bf{a}}(\theta _{ - 90} )} &  \cdots  & {{\bf{a}}(\theta _{ - b - 1} )} & {{\bf{a}}(\theta _{ + b + 1} )} &  \cdots  & {{\bf{a}}(\theta _{ + 90} )}  \\
\end{array}} \right]
\end{equation}
$\textbf{A}_M$ and $\textbf{A}_S$ are sub-matrices of the steering
matrix \textbf{A}. $\textbf{A}_M$ is composed of 2\emph{b}+1
steering vectors with the sampled angles in the mainlobe; while
$\textbf{A}_S$ is constituted with the rest of the steering vectors
in \textbf{A}. The product $ {\bf{w}}^H {\bf{A}}_M $ indicates array
gains of the mainlobe in the beam pattern, and $ {\bf{w}}^H
{\bf{A}}_S $ indicates array gains of the sidelobe. The width of the
first block is 2\emph{b}+1 corresponding to the mainlobe; and the
other block's width is corresponding to the sidelobe. $\gamma _2$ is
the weighting factor that controls the tradeoff between the minimum
variance constraint on the total array output energy and the mixed
norm constraint on the beam pattern to shape the beam pattern,
\emph{b} is an integer representing the bounds of the mainlobe
block. The minimization of $ \left\| {{\bf{w}}^H {\bf{A}}_M }
\right\|_\infty   + \left\| {{\bf{w}}^H {\bf{A}}_S } \right\|_1 $ is
the mixed norm constraint.

As the mixed norm constraint are used in (\ref{eq7}), we name it as
the mixed norm beamformer (MNB). To make a difference, the
beamformer (\ref{eq4}) which was proposed in \cite{sbf} is named as
the standard sparse beamformer. Since the objective function of MNB
is convex, the optimal $ {\bf{w}}_{MNB} $ can be solved out by cvx
\cite{cvx} and SeDuMi \cite{SeDuMi}.

In the mixed norm beamformer (\ref{eq7}), the term $ %
\left\| {{\bf{w}}^H {\bf{A}}_S } \right\|_1 $ enforces the sparse
distribution of the array gains in the sidelobe. That is to say, the
number of the non-trivial array gains is much less than trivial
gains'. And the term $ \left\| {{\bf{w}}^H {\bf{A}}_M }
\right\|_\infty $ lets the array gains in the mainlobe be dense. And
that is to say, the amplitude difference of the array gains in this
area is little. Then the obtained beam pattern would be like this:
most of the non-trivial elements (array gains) are in the mainlobe;
and the rest trivial elements are in the sidelobe. Thus, with most
of entries in the mainlobe being non-trivial, the angle mismatch
would not seriously degenerate the performance; and with most of
entries in the sidelobe being trivial, the effect of interferences
and background noise would decrease.

To illustrate the why the $ \ell _1 $ norm encourage sparse
distribution and the $ \ell _\infty $ norm encourage dense
distribution, a simple geometry is given in Fig. 1. In the figure
the minimization of the $ \ell _1 $ norm of a two dimensional
vector, the $ \ell _2 $ norm of a two dimensional vector and the $
\ell _\infty $ norm of a two dimensional vector are represented as a
smallest rhombus, a smallest circular and a smallest exact square,
respectively. The distortionless constraint can be represented as a
line. The optimal solution for the minimum norm $ \left\|
{{\bf{w}}^H {\bf{A}}} \right\|_p $, ( $ p = 1,2, + \infty $) subject
to the distortionless constraint $ {\bf{w}}^H {\bf{a}}(\theta _0 ) =
1 $ would be the tangent point where the line and the curve
(rhombus, circular, square) meet. Therefore, for the $ \ell _1 $
norm minimization situation, two entries in the solution would has
very different absolute values with high probability; and for the $
\ell _\infty $ norm minimization situation, two entries in the
solution would has considerable similar absolute values with high
probability.

The proposed MNB (\ref{eq7}) fits the beam pattern better than the
standard sparse beamformer in practice, and the performance of the
proposed MNB would be improved.

\section{Simulation Results}

In the simulations, a ULA with 8 half-wavelength spaced antennas is
considered. The AWGN at each sensor is assumed spatially
uncorrelated. The DOA of the SOI is set to be 0$ ^ \circ $, and the
DOAs of three interfering signals are set to be  -30$ ^ \circ $, 30$
^ \circ $, and 70$ ^ \circ $, respectively. The signal to noise
ratio (SNR) is set to be 10 dB, and the interference to noise ratios
(INRs) are assumed to be 20 dB, 20 dB, and 40 dB in  -30$ ^ \circ $,
30$ ^ \circ $, and 70$ ^ \circ $, respectively. 100 snapshots are
used for each simulation. Without loss of generality, \emph{p} is
set to be 1; \emph{b} is set to be 23; and $ \gamma _1 $,  $ \gamma
_2 $ are all set to be 10. The matrix \textbf{A} consists of all
steering vectors in the DOA range of [-90$ ^ \circ $, 90$ ^ \circ $]
with the sampling interval of 1$ ^ \circ $.

To quantify the performance enhancement by the mixed norm
constraint. The SINR is calculated via the following formula:

\begin{equation}
\label{eq10} SINR(b) = \frac{{\sigma _s^2 {\bf{w}} (b)^H
{\bf{a}}(\theta _0 ){\bf{a}}^H (\theta _0 ){\bf{w}} (b)}}{{{\bf{w}}
(b)^H \left( {\sum\limits_{j = 1}^J {\sigma _j^2 {\bf{a}}(\theta _j
){\bf{a}}^H (\theta _j )}  + {\bf{Q}}} \right){\bf{w}} (b)}}
\end{equation}
where $ \sigma _s $  and $ \sigma _j $ are the variances of the SOI
and \emph{j}-th interference, \textbf{Q} is a diagonal matrix with
the diagonal elements being the noise's variances. For a fixed
\emph{b}, $ {\bf{w}}_{IBSB} (b) $ can be obtained via (\ref{eq7}).
Then SINR(b) can be obtained via (\ref{eq10}). Simulations show that
that from \emph{b} = 1 to \emph{b} = 35, The value of SINR increases
gradually at the beginning, and achieves the maximum SINR at $
b_{opt} $ = 23. Then it drops afterwards. We can see that from
\emph{b} = 0 to \emph{b} = $ b_{opt} $, the SINR increases gradually
with the increase of block-sparse constraint strength, and when $
b_{opt} $, the block-sparse constraint starts to mismatch the beam
pattern, and the SINR drops.

Fig. 2 shows beam patterns of the MVDR beamformer (\ref{eq3}), the
standard sparse beamformer (\ref{eq4}), and the MNB (\ref{eq7}) of
1000 Monte Carlo simulations. It is obvious that the best sidelobe
suppression performance is achieved by the MNB (\ref{eq7}). Among
the three beamformers, the MNB (\ref{eq7}) has the lowest array gain
level in sidelobe area, and provides the deepest nulls in the
directions of interference, i.e.,  -30$ ^ \circ $, 30$ ^ \circ $ and
70$ ^ \circ $. The average received SINR by the MVDR beamformer
(\ref{eq3}), the standard sparse beamformer (\ref{eq4}) and the MNB
(\ref{eq7}) are 1.2464 dB, 4.6289 dB and 5.8712 dB.

Fig. 3 shows beam patterns of the beamformers that we have
discussed, with each beamformer having a 4$ ^ \circ $ mismatch
between the steering angle and the DOA of the SOI \cite{doam}. We
can see that the MVDR beamformer has a deep notch in 4$ ^ \circ $,
which is the DOA of the SOI. It can be explained by using the fact
that the MVDR beamformer is designed to minimize the total array
output energy subject to a distortionless constraint in the DOA of
the SOI, so when the steering angle is in 4$ ^ \circ $, instead of
0$ ^ \circ $, the MVDR beam pattern maintains distortionless in 0$ ^
\circ $ while resulting in a deep null in 4$ ^ \circ $. This
observation shows the high sensitivity of the MVDR beamformer to
steering angle mismatch. Comparing beam patterns of beamformers
defined in (\ref{eq4}) and (\ref{eq7}), we can see that the MNB
(\ref{eq7}) further suppresses sidelobe levels and deepens the nulls
for interference avoidance, and has almost the same robustness
against mismatch. In the case of 4$ ^ \circ $ mismatch, the average
received SINR by the MVDR beamformer (\ref{eq3}), standard sparse
beamformer (\ref{eq4}) and the MNB (\ref{eq7}) are 0.0005 dB, 2.0163
dB and 3.2015 dB respectively.

When \emph{b} = 0, the MNB (\ref{eq7}) changes into the standard
sparse beamformer (\ref{eq4}). For $ b \ne 0 $, the beam shaping
property of the beam pattern is exploited. Our simulations also
shows that the optimal block width \emph{b} is between 21 to 27,
when the directions of interferences are in [-90$ ^ \circ $, -15$ ^
\circ $) and [15$ ^ \circ $, 90$ ^ \circ $).

Thus, our proposed beamformer provides improvements in terms of
sidelobe suppression, nulling for interference avoidance, while
maintaining the robustness against the DOA estimation errors, with
respect to existing beamformers.

\section{Conclusion}
The proposed MNB shows superiority to the MVDR beamformer and the
standard sparse beamformer. It outperforms in terms of sidelobe
suppression, nulling for interference avoidance, while maintaining
the robustness against DOA mismatch.

\section*{Acknowledgment}

The authors would like to thank the anonymous reviewers.

This work was supported in part by the National Natural Science Foundation of China under grant 60772146,
the National High Technology Research and Development Program of China (863 Program) under grant 2008AA12Z306
and in part by Science Foundation of Ministry of Education of China under grant 109139.

\ifCLASSOPTIONcaptionsoff
  \newpage
\fi

\begin{figure}[!h]
 \centering
 \includegraphics[scale = 0.47]{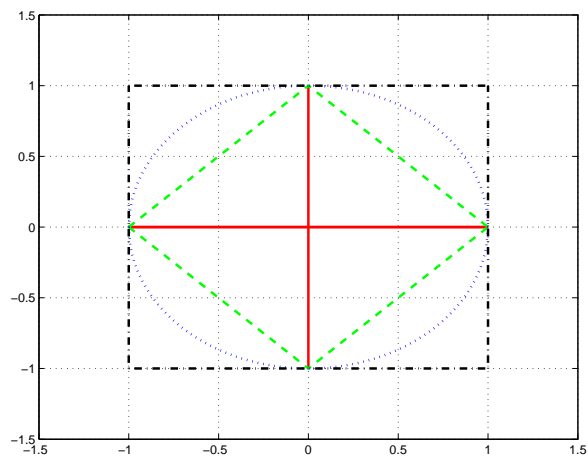}
 \caption{The geometry demonstration of the $ \ell _1 $ norm, $ \ell
_2 $ norm and $ \ell _\infty $ norm in the place.}
 \label{figure0}
\end{figure}

\begin{figure}[!h]
 \centering
 \includegraphics[scale = 0.47]{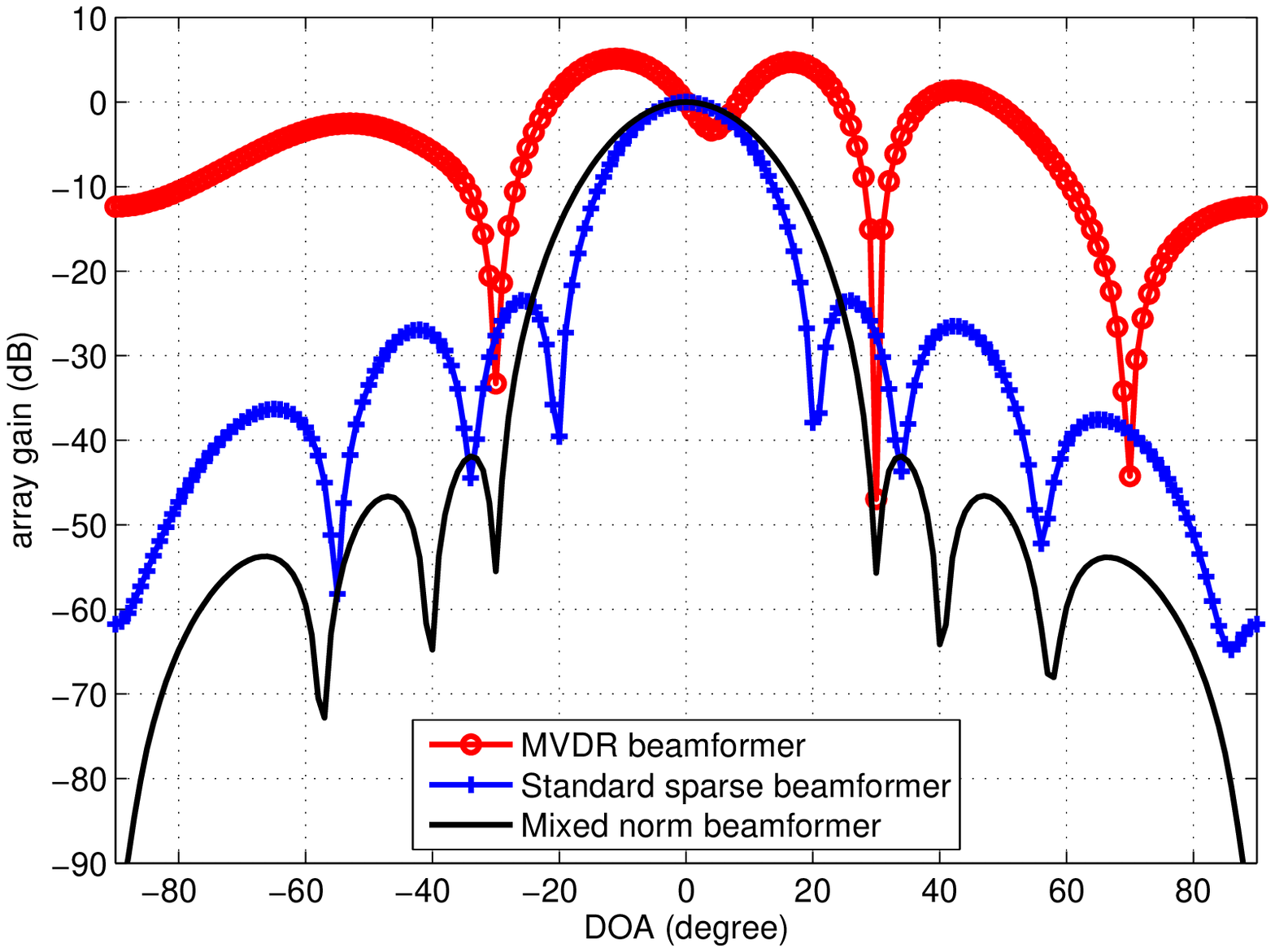}
 \caption{Normalized beam patterns of the MVDR beamformer,
 standard sparse beamformer and the improved block-sparse beamformer,
 without mismatch between the steering angle and the DOA of the SOI.}
 \label{figure1}
\end{figure}

\begin{figure}[!h]
 \centering
 \includegraphics[scale = 0.47]{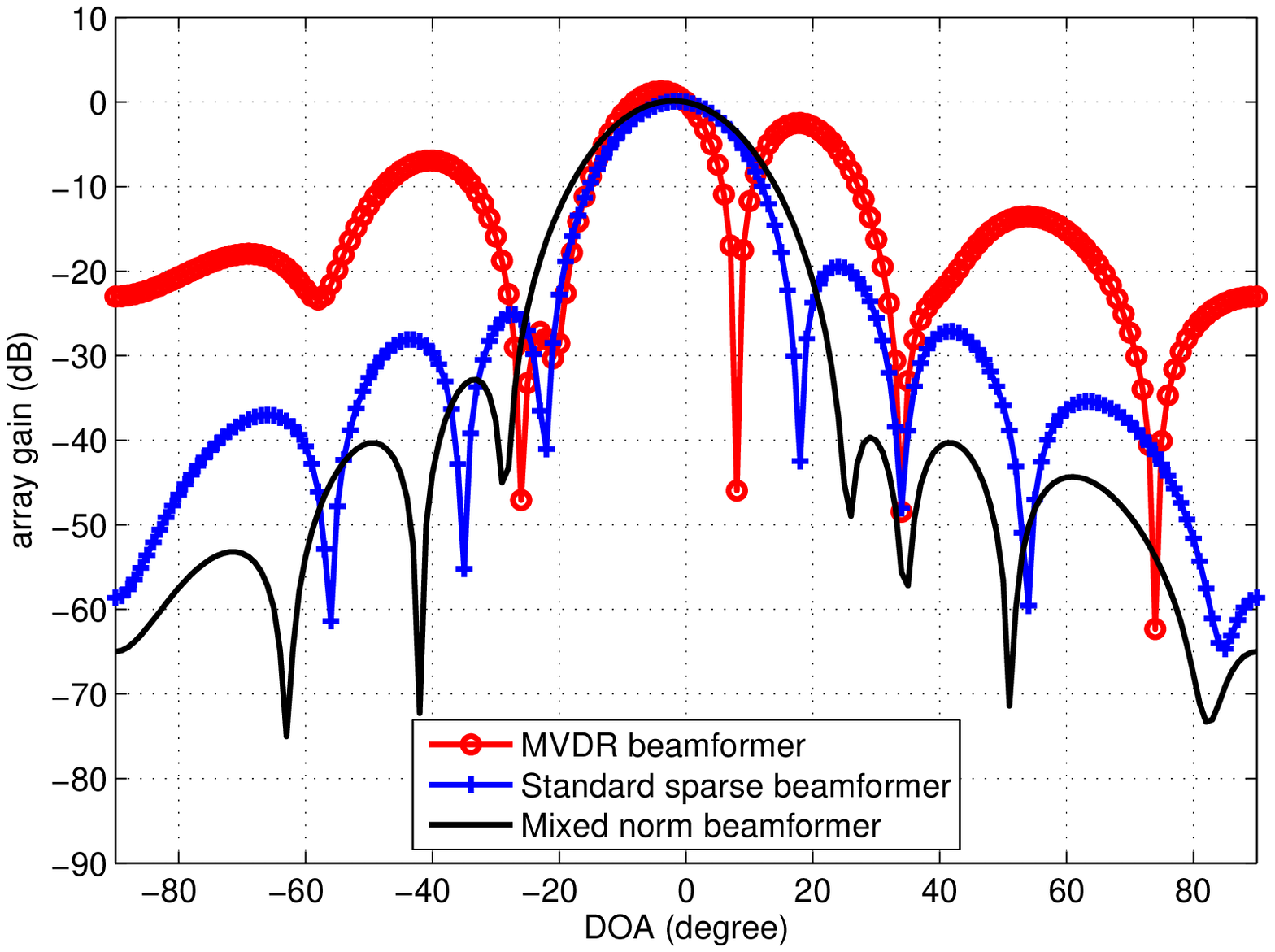}
 \caption{ Normalized beam patterns of the MVDR beamformer,
 standard sparse beamformer and the improved block-sparse beamformer,
 with 4бу mismatch between the steering angle and the DOA of the SOI.}
 \label{figure2}
\end{figure}


\begin{thebibliography}{1}

\bibitem{rab}
 J.~Li and P.~Stoica, \emph{Robust adaptive
beamforming}.\hskip 1em plus  0.5em minus 0.4em\relax New York:
Wiley, 2006.

\bibitem{pamvdr}
M.~Wax and Y.~Anu, "Performance analysis of the minimum variance
beamformer," \emph{IEEE Transactions on Signal Processing}, vol.44,
no.4, pp.928-937, April 1996.


\bibitem{dcrcb}

J.~Li, P.~Stoica, and Z.~Wang, "Doubly constrained robust capon
beamformer," \emph{IEEE Transactions on Signal Processing}, Vol. 52,
pp. 2407-2423, 2004.

\bibitem{lqc}

L.~Zhang, W.~Liu, and R.~J.~Langley, "A minimum variance beamformer
with linear and quadratic constraints based on uniform linear
antenna arrays," \emph{Antennas and Propagation Conference 2009
(LAPC 2009)}, Loughborough, Nov. 16-17, 2009, pp. 585-588.

\bibitem{rbf}

R.~Savitha, S.~Vigneswaran, S.~Suresh, and N.~Sundararajan,
"Adaptive beamforming using complex-valued radial basis function
neural networks," \emph{2009 IEEE Region 10 Conference (TENCON
2009)}, Singapore, Jan. 23-26, 2009, pp. 1-6.




\bibitem{sbf}
Y.~Zhang, B.~P.~Ng, and Q.~Wan, "Sidelobe suppression for adaptive
beamforming with sparse constraint on beam pattern,"
\emph{Electronics Letters}, vol. 44, iss. 10, pp. 615-616, May 2008.

\bibitem{bdrao}
B.~D.~Rao, K.~Engan, S.~F.~Cotter, J.~Palmer, and K.~K.~Delgado,
"Subset selection in noise based on diversity measure minimization,"
\emph{IEEE Transactions on Signal Processing}, vol.51, no.3,
pp.760-770, March 2003.

\bibitem{bp}
S.~Chen and D.~Donoho, "Basis pursuit," \emph{28th Asilomar
Conference on Signals, Systems and Computers}, vol. 1, pp. 41-44,
October-November 1994.


\bibitem{doam}
C.~Chen and P.~P.~Vaidyanathan, "Quadratically constrained
beamforming robust against direction-of-arrival mismatch,"
\emph{IEEE Transactions on Signal Processing}, vol. 55, iss. 8, pp.
4139 - 4150, August 2007.
%
%

\bibitem{bs}

M.~Stojnic,~F.~Parvaresh,~and B.~Hassibi, "On the reconstruction of
block-sparse signals with an optimal number of measurements,"
\emph{IEEE Transaction on Signal Processing}, Vol. 57, No. 8, pp.
3075-3085, August 2009.

\bibitem{cvx}

M.~Grant,~S.~Boyd,~Y.~Ye, cvx user' guide for cvx version
1.1,~[online] accessible at :
http://www.stanford.edu/~boyd/index.html

\bibitem{SeDuMi}

J.~Sturm,~"Using sedumi 1.02, a matlab toolbox for optimization over
symmetric cones," \emph{Optimization Methods and Software}, Vol. 11,
No. 12, pp. 625-653, 1999.


\end{thebibliography}
\end{document}